\documentclass[sigconf]{acmart}

\usepackage{booktabs} 
\usepackage{hyperref}
\usepackage{amsmath}
\usepackage{multirow}

\usepackage{multicol}
\usepackage{graphicx,dblfloatfix}
\usepackage{hhline,colortbl}
\usepackage[ruled, linesnumbered]{algorithm2e}
\usepackage{color, colortbl}
\definecolor{LightCyan}{rgb}{0.8,0.8,1}
\definecolor{Maroon}{rgb}{1,0.8,0.8}
\definecolor{Gray}{rgb}{0.75,0.75,0.75}

\newcommand{\camready}[1]{\textcolor{black}{#1}}
\usepackage{booktabs}
\DontPrintSemicolon
\usepackage{flushend}

\usepackage{tikz}
\newcommand*\circled[1]{\tikz[baseline=(char.base)]{
            \node[shape=circle,draw,inner sep=0.75pt] (char) {#1};}}

\AtBeginDocument{%
  \providecommand\BibTeX{{%
    \normalfont B\kern-0.5em{\scshape i\kern-0.25em b}\kern-0.8em\TeX}}}

\begin{document}



\title{Performance Analysis of Priority-Aware NoCs with \\ Deflection Routing under Traffic Congestion}



\author{Sumit K. Mandal$^1$, Anish Krishnakumar$^1$, Raid Ayoub$^2$, Michael Kishinevsky$^2$, Umit Y. Ogras$^1$}
\affiliation{
\institution{$^1$Dept. of ECE, University of Wisconsin-Madison; 
$^2$Intel Corporation, Hillsboro, OR
}
}

\vspace{-2mm}
\begin{abstract}
%

Priority-aware networks-on-chip (NoCs) are used in industry to achieve predictable latency under different workload conditions.
These NoCs incorporate deflection routing to minimize queuing resources within routers and achieve low latency during low traffic load.
However, deflected packets can exacerbate congestion during high traffic load since they consume the NoC bandwidth. 
State-of-the-art analytical models for priority-aware NoCs ignore deflected traffic despite its significant latency impact during congestion. 
This paper proposes a novel analytical approach to estimate end-to-end latency of \textit{priority-aware} NoCs with \textit{deflection routing} under \textit{bursty and heavy traffic} scenarios.
Experimental evaluations show that the proposed technique outperforms alternative approaches and estimates the average latency for real applications with less than 8\% error compared to cycle-accurate simulations.

\end{abstract}


\thanks{This paper has been published in the proceedings of ICCAD 2020.
	
This work was supported partially by Strategic CAD Labs, Intel Corporation, USA.

Author's addresses: S. K. Mandal, A. Krishnakumar {and} U. Y. Ogras, Department of
Electrical and Computer Engineering, University of Wisconsin-Madison, WI,
53706.   \newline 
Emails: \{skmandal, anish.n.krishnakumar, uogras\}@wisc.edu;

R. Ayoub {and} M. Kishinevsky, Intel Corporation, 2111 NE 25th Ave.,
Hillsboro, OR 97124; emails: \{raid.ayoub, michael.kishinevsky\}@intel.com

Link to the short presentation: \url{https://www.youtube.com/watch?v=rMZWC8imO8k&ab_channel=SumitKumarMandal}

Link to the full presentation:
\url{https://www.youtube.com/watch?v=OHqwSLxR2UU&ab_channel=SumitKumarMandal}
}

\settopmatter{printacmref=false}
\pagestyle{empty}


\maketitle

\vspace{-4mm}
\section{Introduction}


Pre-silicon design-space exploration and system-level simulations constitute a crucial component of the industrial design cycle~\cite{palesi2002multi,ghosh2003analytical}.
They are used to confirm that new generation designs meet power-performance targets before labor- and time-intensive RTL implementation starts~\cite{Binkert2011Gem5}. 
Furthermore, virtual platforms combine power-performance simulators and functional models to enable firmware and software development while hardware design is in progress~\cite{leupers2011virtual}. 
These pre-silicon evaluation environments incorporate cycle-accurate NoC simulators due to the criticality of shared communication and memory resources in overall performance~\cite{agarwal2009garnet, jiang2013detailed}.
However, slow cycle-accurate simulators have become the major bottleneck of pre-silicon evaluation. 
Similarly, exhaustive design-space exploration is not feasible due to the long simulation times. 
Therefore, there is a strong need for fast, yet accurate, analytical models to replace cycle-accurate simulations to increase the speed and scope of pre-silicon evaluations~\cite{yoo2002automatic}.

\begin{figure}[t]
	\centering
	\includegraphics[width=0.73\linewidth]{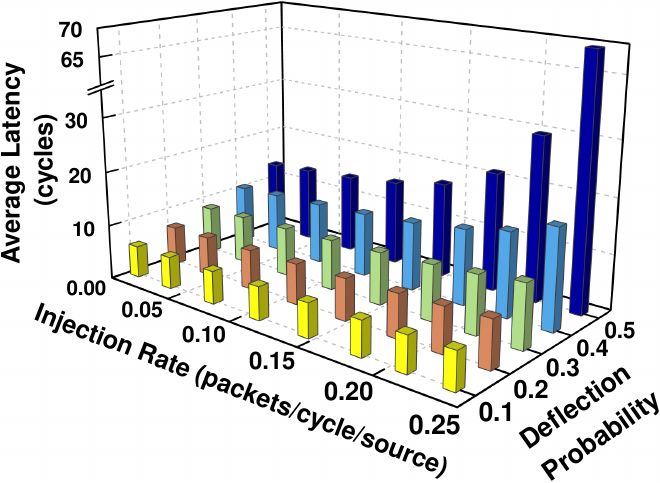}
	\vspace{-3mm}
	\caption{Cycle-accurate simulations on a 6$\times$6 NoC show that the average latency increases significantly with larger deflection probability ($p_d$) at the sink.}
	\label{fig:motivation}
	\vspace{-5mm}
\end{figure}

Analytical NoC performance models are used primarily for fast design space exploration since they provide significant speed-up compared to detailed simulators~\cite{ogras2010analytical, kiasari2013analytical, kashif2014bounding, qian2015support}.
However, most existing analytical models fail to capture two important aspects of industrial NoCs~\cite{jeffers2016intel}.
\textit{First,} they do not model routers that employ priority arbitration. 
\textit{Second,} existing analytical models assume that the destination nodes always sink the incoming packets. 
In reality, network interfaces between the routers and cores have finite (and typically limited) ingress buffers. Hence, packets bounce (i.e., they are deflected) at the destination nodes 
when the ingress queue is full.
\camready{Recently proposed performance models target priority-aware NoCs~\cite{mandal2019analytical,mandal2020analytical}}.
However, these ignore deflection due to finite buffers 
and uses the packet injection rate as the primary input. 
This is a significant limitation 
since the deflection probability ($p_d$) increases both the hop count and 
traffic congestion. 
Indeed, Figure~\ref{fig:motivation} shows that the average NoC latency increases significantly with the probability of deflection. 
\camready{For example, the average latency varies from 6--70 cycles for an injection rate of 0.25 packets/cycle/source when $p_d$ varies from 0.1--0.5.}
\camready{Therefore, performance models for priority-aware NoCs have to account for deflection probability at the destinations.}

This work proposes an accurate analytical model for \textit{priority-aware NoCs with deflection routing under bursty traffic}.
In addition to increasing the hop count, 
deflection routing also aggravates traffic congestion due to extra packets traveling in the network. 
Since the deflected
packets also have a complex effect on the egress queues of the traffic sources, analytical modeling of priority-aware NoCs with deflection routing is challenging.
To address this problem, we first need to approximate the probability distribution of inter-arrival time of deflected packets.
Specifically, we compute the first two moments of inter-arrival time of deflected packets since we consider bursty traffic. To this end, the proposed approach starts with a canonical queuing system with deflection routing.
We first model the distribution of deflected traffic and the average queuing delay for this system.
However, this methodology is not scalable when the network has multiple queues with complex interactions between them. 
Therefore, we also propose a superposition-based technique to obtain the waiting time of the packets in
\camready{arbitrarily}
sized industrial NoCs. 
This technique decomposes the queuing system into multiple subsystems.
The structure of these subsystems is similar to the canonical queuing system.
After deriving the analytical expressions for the parameters of the distribution model of deflected packets
of individual subsystems, we superimpose the result to solve the original system with multiple queues. 
Thorough experimental evaluations with industrial NoCs and their cycle-accurate simulation models show that the proposed technique significantly outperforms prior approaches~\cite{kiasari2013analytical, mandal2019analytical}. 
In particular, the proposed technique achieves less than 8\% modeling error when tested with real applications from different benchmark suites. 
%
\textit{The major contributions of 
\camready{this} work are as follows}:
\vspace{-1mm}
\begin{itemize}
    \item An accurate performance model for priority-aware NoCs with deflection routing under bursty traffic,
    \item An algorithm to obtain end-to-end latency using the proposed performance model,  
    \item Detailed experimental evaluation with industrial priority-aware NoC under varying degrees of deflection.
\end{itemize}

The rest of this paper is organized as follows.
Section~\ref{sec:rel_work} summarizes the related research.
Section~\ref{sec:background} presents the background and overview 
\camready{of}
the proposed work.
Section~\ref{sec:method} describes the proposed methodology to construct the analytical model for priority-aware NoCs\camready{,} which considers deflection routing.
Section~\ref{sec:expt_eval} details experimental evaluations, and 
Section~\ref{sec:concl} concludes the paper.

\section{Related Work} \label{sec:rel_work}



Deflection routing was first introduced in the domain of optical NoC as hot-potato routing~\cite{borodin1997deterministic}.
Later, it was adapted for the NoCs used in high-performance SoCs to minimize buffer requirements and increase energy efficiency~\cite{moscibroda2009case, fallin2011chipper, fallin2012minbd}.
This routing mechanism always assigns the packets to a free output port of a router, even if the assignment does not result in minimum latency.
This way, the buffer size requirement in the routers is minimized.
Authors in~\cite{lu2006evaluation} perform a thorough study on the effectiveness of deflection routing for different NoC topology and routing algorithm.
Deflection routing is also used in industrial priority-aware NoC~\cite{jeffers2016intel}. 
Since arbitrary deflections can cause livelocks and unpredictable latency, industrial priority-aware NoCs deflect the packets only at the destination nodes when the ingress buffer is full. Furthermore, the deflected packets always remain within the same row or column, and they are guaranteed to be sunk after a fixed number of deflections.

NoC performance analysis techniques have been used for design space exploration and architectural studies such as buffer sizing~\cite{ogras2010analytical,wu2010analytical, petracca2009photonic}.
However, most of these techniques do not consider NoCs with priority arbitration and deflection routing, which are the key features of industrial NoCs~\cite{jeffers2016intel}. 
Performance analysis of priority-aware queuing networks has also been studied for off-chip networks~\cite{bertsekas1992data, bolch2006queueing, walraevens2004discrete}.
These analytical models consider the queuing networks in continuous time.
However, each transaction in NoC happens at each clock cycle.
Therefore, the underlying queuing system needs to be considered in the discrete time domain.
A performance analysis technique for a priority-aware queuing network in discrete time domain is presented in ~\cite{walraevens2004discrete}.
However, this technique suffers from high complexity for a complex queuing network, hence not applicable to industrial priority-aware NoCs.

A recent technique targets priority-aware NoCs~\cite{kiasari2013analytical}, but it considers only a single class of packets in each queue of the network.
In contrast, industrial priority-aware NoCs have multiple classes of packets that can exist in the same queue.
NoCs with multiple priority traffic classes 
\camready{has}
recently been analyzed in~\cite{mandal2019analytical}.
However, this analysis assumes that the input traffic follows a geometric distribution. 
This technique has limited applications since industrial NoCs can experience bursty traffic. 
Furthermore, it does not consider deflection routing.
Since deflection routing increases traffic congestion, it is crucial to incorporate this aspect while constructing performance models.
An analytical bound on maximum delay in networks with deflection routing is presented in~\cite{brassil1995bounds}.
However, evaluating maximum delay is not useful since it leads to significant overestimation.
Another analytical model for NoCs with deflection routing is proposed in~\cite{ghosh2010analytical}.
The authors first compute the blocking probability at each port of a router using an M/G/1 queuing model.
Then, they compute the contention matrix at each router port.
The average waiting time of packets at each port is computed using the contention matrix.
However, this analysis ignores different priority classes and applies to only continuous-time queuing systems.



In contrast to prior work, we propose a performance analysis technique that considers \textbf{\textit{both priority-aware NoCs with deflection routing under bursty and high traffic load.}}
The proposed technique applies the superposition principle to obtain the statistical distribution of the deflected packets. 
Using this distribution, it computes the average waiting time for each queue. 
To the best of our knowledge, this is the first analytical model for priority-aware industrial NoCs with deflection routing under high traffic load.
\section{Background and Overview} \label{sec:background}

\begin{figure}[b]
	\centering
	\vspace{-4mm}
 	\includegraphics[width=1.0\linewidth]{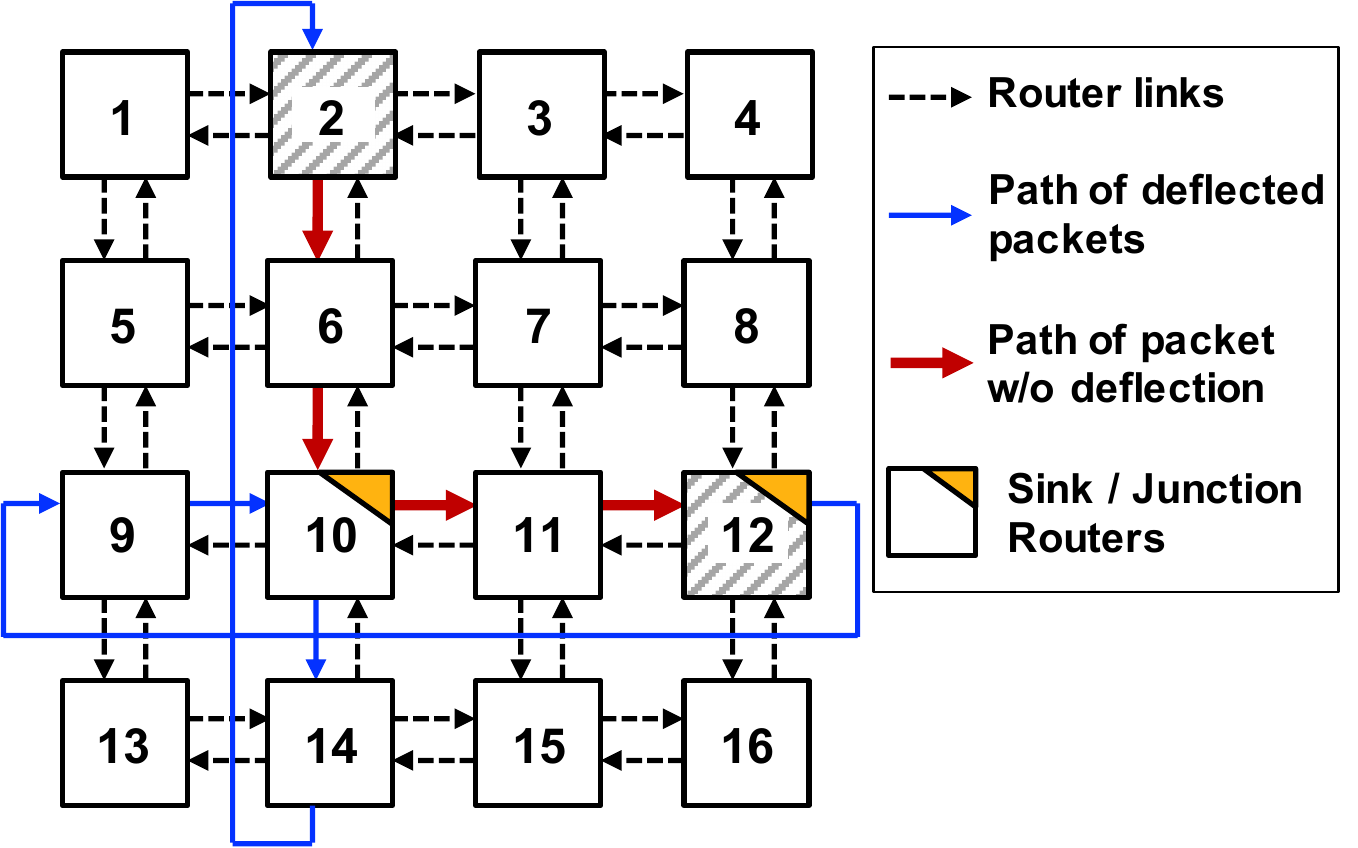}
	\caption{A representative 4$\times$4 mesh with deflection routing.}
	\label{fig:bnc_expl}
\end{figure}

\begin{figure*}[!b]
	\centering
 	\includegraphics[width=0.85\linewidth]{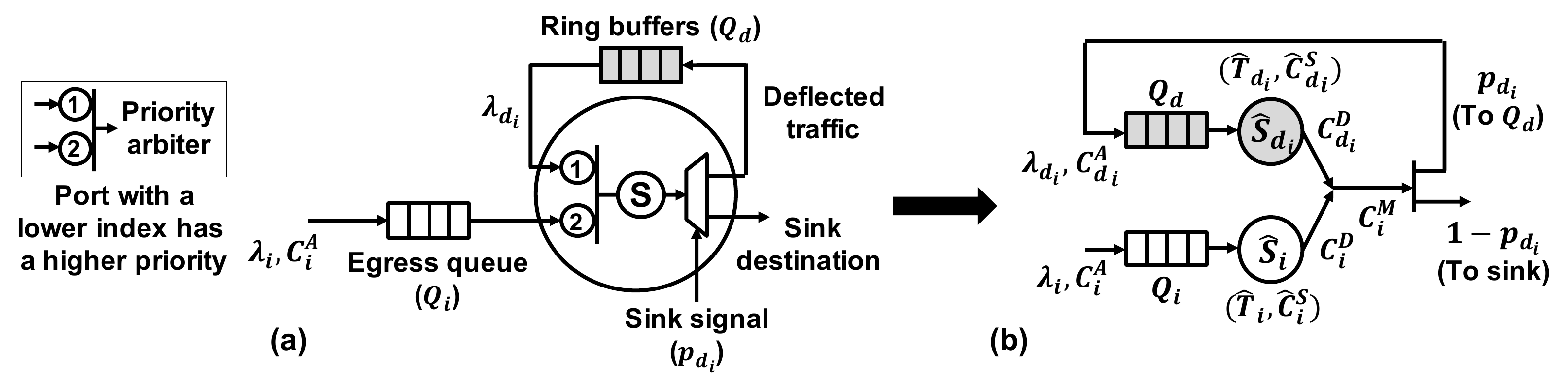}
	\caption{(a) Queuing system of a single class with deflection routing (b) 
	Approximate queuing system to compute $C_{d_i}^A$.}
	\vspace{-15mm}
	\label{fig:canon_struct}
\end{figure*}

\subsection{Assumptions and Notations}
\noindent\textbf{Architecture:}
%
This work considers priority-aware NoCs used in 
high-end servers and many core architectures~\cite{jeffers2016intel}.
Each column of the NoC architecture, shown in Figure~\ref{fig:bnc_expl}, is also used in client systems  such as Intel i7 processors~\cite{rotem2015intel}.
Hence, the proposed analysis technique is broadly applicable to a wide range of industrial NoCs.

In priority-aware NoCs, the packets already in the network have higher priority than the packets waiting in the egress queues of the sources.
Assume that Node 2 in Figure~\ref{fig:bnc_expl} sends a packet to Node 12 following Y-X routing (highlighted by red arrows). Suppose that a packet in the egress queue of Node 6 collides with this packet.
The packet from Node 2 to Node 12 will take precedence since the packets already in the NoC have higher priority. 
Hence, packets experience a queuing delay at the egress queues 
but have predictable latency until they reach the destination or turning point (Node 10 in Figure~\ref{fig:bnc_expl}). Then, it competes with the packets already in the corresponding row. 
That is, the path from the source (Node 2) to the destination (Node 12) can be considered as two segments\camready{,} which consist of a queuing delay followed by a predictable latency. 

Deflection in priority-aware NoCs happens when the ingress queue at the turning point (Node 10) or final destination (Node 12) become  full. This can happen if the receiving node, such as a cache controller, cannot process the packets fast enough. The probability of observing a full queue increases with smaller queues (needed to save area) and heavy traffic load from the cores. 
If the packet is deflected at the destination node, it circulates within the same row, as shown in Figure~\ref{fig:bnc_expl}.
Consequently, a combination of regular and deflected traffic can load the corresponding row and pressure the ingress queue at the turning point (Node 10). This, in turn, can lead to deflection on the column and propagates the congestion towards the source. 
Finally, if a packet is deflected more than a specific number of times, it reserves a slot in the ingress queue. This bounds the maximum number of deflections and avoids livelock.

\noindent\textbf{Traffic:}
%
Industrial priority-aware NoCs can experience bursty traffic, which is characteristic of real applications~\cite{bogdan2010workload, qian2015support}. 
This work considers generalized geometric (GGeo) distribution for the input traffic, which takes burstiness into account~\cite{kouvatsos1994entropy}.
GGeo traffic is characterized by an average injection rate ($\lambda$) and the coefficient of variation of inter-arrival time ($C^A$).
We define a traffic class as the traffic 
\camready{of}
each source-destination pair.
The average injection rate and coefficient of variation of inter-arrival time of class-$i$ are denoted by $\lambda_i$ and as $C_{i}^A$ respectively, as shown in Table~\ref{tab:symbols_used}.
Finally, the mean service time and coefficient of variation of inter-departure time of class-$i$ are denoted as $T_i$ and $C_{i}^S$.

\subsection{Overview of the Proposed Approach} \label{sec:overview}

Our goal is to construct an accurate analytical model to compute the end-to-end latency for priority-aware NoCs with deflection routing.
The proposed approach can be used to accelerate full system simulations and also to perform design space exploration.
We assume that the parameters of the GGeo distribution of the input traffic to the NoC ($\lambda, C^A$) are known from the knowledge of the application.
The proposed model uses the deflection probability ($p_d$) as the second major input, in contrast to existing techniques that ignore deflection.
Its range is found from architecture simulations as a function of the NoC architecture (topology, number of processors, and buffer sizes). Its analytical modeling is left for future work. 
The proposed analytical model utilizes the distribution of the input traffic to the NoC ($\lambda, C^A$) and the deflection probability ($p_d$) to compute the average end-to-end latency as a sum of four components: (1) Queuing delay at the source, (2) the latency from the source router to the junction router, (3) queuing delay at the junction router, and (4) the latency from the junction router to the destination. Note that all these components account for deflection\camready{,} and it is challenging to compute them, especially under high traffic load.

\begin{table}[t]
\caption{Summary of the notations used in this paper.}
\centering
	\vspace{-3mm}
	\label{tab:symbols_used}
	\begin{tabular}{@{}ll@{}}
\hline
$\lambda_i$     & Arrival rate of class-$i$                                                                                                    \\ \hline
$p_{d_j}$       & Deflection probability at sink-$j$                                                                                                    \\ \hline
$T_i$, $\widehat{T}_i$  & Original and modified mean service time of class-$i$                                                                                                    \\ \hline
                                      
$\rho_i$        & Mean server utilization of class-$i$ (=$\lambda_i T_i$)                                                                             \\ \hline
$C_{i}^A$  & \begin{tabular}[c]{@{}l@{}} Coefficient of variation \\ of inter-arrival time of class-$i$  \end{tabular} \\ \hline
$C_{i}^S, \widehat{C}_{i}^S$    & \begin{tabular}[c]{@{}l@{}} Coefficient of variation of original \\ and modified service time of 
class-$i$\end{tabular} \\ \hline
${C}_{i}^D$  & \begin{tabular}[c]{@{}l@{}} Coefficient of variation \\ of inter-departure time of class-$i$\end{tabular} \\ \hline
${C}_{i}^M$  & \begin{tabular}[c]{@{}l@{}} Coefficient of variation of inter-departure time \\ of merged traffic of class-$i$\end{tabular} \\ \hline
$W_i$       & Mean waiting time of class-$i$                                                                                                      \\ \hline
\end{tabular}
\end{table}


\section{Methodology} \label{sec:method}

This section presents the proposed performance analysis technique for estimating the end-to-end latency for priority-aware NoCs with deflection routing.
We first construct a model for a canonical system with a single traffic class,  where the deflected traffic distribution is approximated using a GGeo distribution (Section~\ref{sec:canon_struct}).
Subsequently, we introduce a scalable approach for a network with multiple traffic classes. 
In this approach, we first develop a solution for the canonical system. Then, employ the principle of superposition to extend the analytical model to larger and realistic NoCs with multiple traffic classes (Section~\ref{sec:superpos}).
Finally, we propose an algorithm that uses our analytical models to compute the average end-to-end latency for a priority-aware NoC with deflection routing (Section~\ref{sec:algorithm}).

\subsection{An Illustration with a Single Traffic Class} \label{sec:canon_struct}

Figure~\ref{fig:canon_struct}(a) shows an example of a single class input traffic and egress queue that inject traffic to a network with deflection routing. 
The input packets are buffered in the egress queue $Q_i$ (analogous to the packets stored in the egress queue of Node 2 in Figure~\ref{fig:bnc_expl}). 
We denote the traffic of $Q_i$ as class-$i$, which is modeled using GGeo distribution with two parameters ($\lambda_i, C_i^A$).
The packets in $Q_i$ are dispatched to a priority arbiter and assigned a low priority, marked with \circled{2}. \camready{In contrast,} 
the packets already in the network have a high priority\camready{,} which are routed to the port marked with \circled{1}. 
The packet traverses a certain number of hops (similar to the latency from the source router to the junction router in Figure~\ref{fig:bnc_expl}) and reaches the destination.
Since the number of hops is constant for a particular traffic class, \textit{we omit these details in Figure~\ref{fig:canon_struct}(a) for simplicity}.
If the ingress queue at the destination is full
(with probability $p_{d_i}$), the packet is deflected back into the network. Otherwise, it is consumed at the destination (with probability $1-p_{d_i}$). 
Deflected packets travel through the NoC (within the column or row as illustrated in Figure~\ref{fig:bnc_expl}) and pass through the source router, but this time with higher priority. 
The profile of the deflected packets in the network is modeled by a buffer ($Q_{d}$) in Figure~\ref{fig:canon_struct}(a), 
since they remain in order and have a fixed latency from the destination to the original source.
This process continues until the destination 
\camready{can}
consume the deflected packets. 


Our goal is to compute the average waiting time $W_i$ in the source queue, i.e., components 1 and 3 of the end-to-end latency described in Section~\ref{sec:overview}. 
\camready{To}
obtain $W_i$, we first need to derive the analytical expression for the rate of deflected packets of class-$i$ ($\lambda_{d_i}$) and the coefficient of variation of inter-arrival time of the deflected packets ($C_{d_i}^A$) as follows.

\noindent\textbf{Rate of deflected packets ($\lambda_{d_i}$)}: $\lambda_{d_i}$ is obtained by calculating the average number of times a packet is deflected ($N_{d_i}$) until it is consumed at the destination
as:
\begin{align} \nonumber
    N_{d_i} & = p_{d_i}(1-p_{d_i}) + 2p_{d_i}^2(1-p_{d_i}) + \hdots + n p_{d_i}^n(1-p_{d_i}) + \hdots \\
    & = \sum_{n=1}^{\infty} np_{d_i}^n (1-p_{d_i}) = \frac{p_{d_i}}{1-p_{d_i}}
\end{align}
Therefore, $\lambda_{d_i}$ can be expressed as:
\begin{flalign}
    \lambda_{d_i} &= \lambda_i N_{d_i} = \lambda_i \frac{p_{d_i}}{1-p_{d_i}} 
\end{flalign}

\noindent\textbf{Coefficient of variation of inter-arrival time of deflected packets ($C_{d_i}^A$):} To compute $C_{d_i}^A$, 
the priority related interaction between the deflected traffic of $Q_{d}$ and new injections in $Q_i$ must be captured.
This computation is more involved due to the priority arbitration between the packets in $Q_{d}$ and $Q_i$ that involve a 
circular dependency. 
We tackle this problem by transforming the system in Figure~\ref{fig:canon_struct}(a) into an approximate representation shown in Figure~\ref{fig:canon_struct}(b) to simplify the computations. 
The idea here is to transform the priority queuing with a shared resource into separate queue nodes (queue + server) with a modified server process. 
This transformation enables the decomposition of $Q_{d}$ and $Q_i$ and their shared server into individual queue nodes with servers $\widehat{S}_{d}$ and $\widehat{S}_i$ respectively.
The departure traffic from these two nodes 
\camready{merge}
at the destination,
consumed with a probability $1-p_{d_i}$ and 
\camready{deflected}
otherwise. 

The input traffic to the egress queue\camready{,} as well as the deflected traffic\camready{,} may exhibit bursty behavior. 
\camready{Indeed, the deflected traffic distribution can be bursty because of the server-process effect and the priority interactions between the input traffic and the deflected traffic, even when the input traffic is not bursty.}
Therefore, we approximate the distribution of the deflected traffic via GGeo distribution. 
To compute the parameters of the GGeo traffic\camready{,} we need to apply the principle of maximum entropy (ME) as shown in ~\cite{kouvatsos1994entropy}. 
To obtain the modified service process of class-$i$, we first calculate the probability of no packets in $Q_i$ and in its corresponding server (i.e., $p_{Q_i}(0)$) using ME as,

%
\begin{equation} \label{eq:empty_q}
    p_{Q_i}(0) = 1 - \rho_{i} -  \rho_{d_i} \frac{\overline{n}_{i}}{\overline{n}_{i}  + \rho_{i} + \rho_{d_i}}
\end{equation}
%
where $\rho_{i}$ and $\rho_{d_i}$ denote the utilization of the respective servers, and 
$\overline{n}_{i}$ is the occupancy of class-$i$ in $Q_i$. Next, we apply Little's law to compute the first order moment of modified service time 
($\widehat{T}_i$) as:
\begin{equation} \label{eq:modif_t}
    \widehat{T}_i = \frac{1-p_{Q_i}(0)}{\lambda_i}
\end{equation}
Subsequently, we obtain the effective coefficient of variation $\widehat{C}_{i}^S$ as:
\begin{equation} \label{eq:modif_cs2}
    (\widehat{C}_{i}^S)^2 = \frac{(1-\widehat{\rho}_i)(2\overline{n}_i + \widehat{\rho}_i) - \widehat{\rho}_i (C_{i}^A)^2}{\widehat{\rho}_i^2}
\end{equation}
where $\widehat{\rho}_i = \lambda_i \widehat{T}_i$.
We follow similar steps (Equation~\ref{eq:empty_q} -- Equation~\ref{eq:modif_cs2}) for the deflected traffic to obtain $\widehat{T}_{d_i}$ and $\widehat{C}_{d_i}^S$.
With the modified service process, the coefficients of variation of inter-departure time of the packets 
in $Q_{d}$ (${C}_{d_i}^D$) and $Q_i$ (${C}_{i}^D$) are computed using the process merging method~\cite{pujolle1986solution}.
Then, we find the coefficient of variation \camready{(${C}_{i}^M$)}
of the merged traffic from queues $Q_{d}$ and $Q_i$ as:

\vspace{-2mm}
\begin{equation}
    ({C}_{i}^M)^2 = \frac{1}{\lambda_{d_i} + \lambda_i} (\lambda_{d_i}({C}_{d_i}^D)^2 + \lambda_i({C}_{i}^D)^2)
\end{equation}
We note that ${C}_{i}^M$ is a function of the coefficient of variation of the inter-arrival time of deflected traffic $C_{d_i}^A$.
Since part of this merged traffic is consumed at the sink, we apply the traffic splitting method from ~\cite{pujolle1986solution} to approximate $C_{d_i}^A$ as:

\vspace{-3mm}
\begin{equation}
    ({C}_{d_i}^A)^2 = 1 + p_{d_i}(({C}_{i}^M)^2  - 1)
\end{equation}
Finally, we extend the priority-aware formulations in continuous time domain~\cite{bolch2006queueing} to discrete time domain to obtain the average waiting time of the packets in $Q_{d_i}$ and $Q_i$:
\begin{flalign} \label{eq:W_0}
    & W_{d_i} = \frac{\rho_{d_i}(T_{d_i}-1) + \rho_i(T_i-1) + T_{d_i}(({C}_{d_i}^A)^2 + \lambda_{d_i} - 1)}{2(1-\rho_{d_i})} \\ \label{eq:W_1}
    & W_i = \frac{\rho_{d_i}(T_{d_i}+1) + 2\rho_{d_i} W_{d_i} + \rho_i(T_i-1) + T_i((C_{i}^A)^2 + \lambda_i - 1)}{2(1-\rho_i - \rho_{d_i})}
\end{flalign}

\begin{figure*}[t]
	\centering
	\vspace{-15mm}
 	\includegraphics[width=1\linewidth]{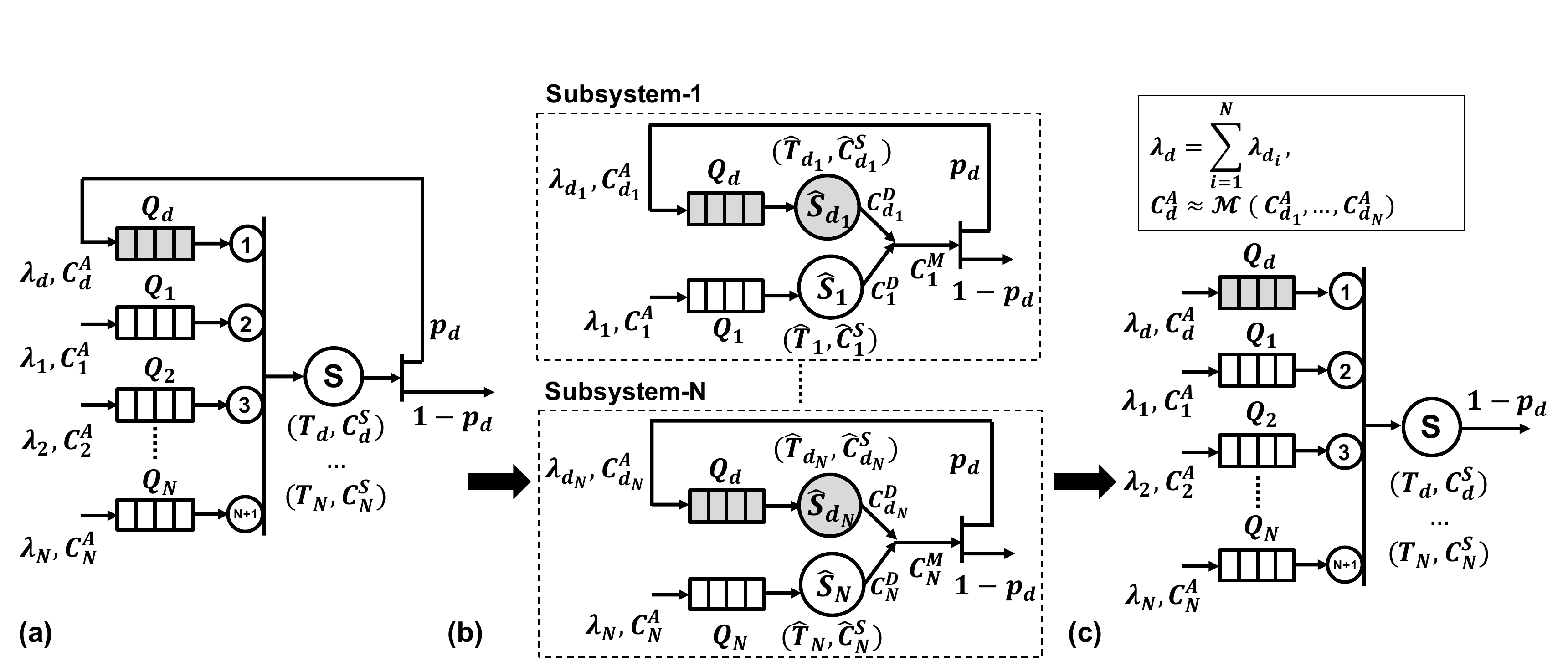}
	\caption{(a) Queuing system with $N$ classes with deflection routing, (b) Decomposition into $N$ subsystems to calculate GGeo parameters of deflected traffic per class, (c) Applying superposition to obtain the GGeo parameters of overall deflected traffic. $\mathcal{M}$ denotes the merging process.}
	\vspace{-5mm}
	\label{fig:superpos}
\end{figure*}

\vspace{-4mm}
\subsection{\hspace{-3mm}Queuing System with Multiple Traffic Classes} \label{sec:superpos}

The analytical model for the system with a single class presented in Section~\ref{sec:canon_struct} becomes intractable with a higher number of traffic classes. 
This section introduces a scalable approach based on the superposition principle
that builds upon our canonical system used in Section~\ref{sec:canon_struct}.

Figure~\ref{fig:superpos}(a) shows an example with priority arbitration and $N$ egress queues, one for each traffic class. 
\textit{We note that this queuing system is a simplified representation of a real system}.
The packets routed to port~\circled{$i$} have higher priority than those routed to port~\circled{$j$} for $i<j$.
The deflected traffic in the network is buffered in $Q_d$, which has the highest priority in the queuing system.
%
%
The primary goal is to model the queuing time of the packets of each traffic class. 
Modeling the coefficient of variations of the deflected traffic becomes harder since deflected packets interact with all traffic classes rather than a single class. These interactions complicate the analytical expressions significantly.

Priority arbitration enables us to sort the queues in the order at which the packets are served.
The queue of the deflected packets has the highest priority, while the rest are ordered with respect to their indices.
Due to this inherent order between the priority classes, their impact on the deflected traffic distribution can be approximated as being independent of each other. This property enables us to decompose the queuing system into multiple subsystems and model each subsystem separately, 
as illustrated in Figure~\ref{fig:superpos}(b).  
Then, we apply the principle of superposition to obtain the parameters of the GGeo distribution of the deflected traffic. 
Note that \textit{each of these subsystems is identical to the canonical system analyzed in Section~\ref{sec:canon_struct}}. 
Hence, we first compute $\lambda_{d_i}$ and $C_{d_i}^A$ of each subsystem-$i$ following the procedure described in Section~\ref{sec:canon_struct}. Subsequently, we apply the superposition principle to $\lambda_{d_i}$ and $C_{d_i}^A$ for $i = 1 \hdots N$ to obtain the GGeo distribution parameters of the deflected traffic ($\lambda_d, C_d^A$).

In general, we obtain the GGeo distribution parameters of the deflected traffic corresponding to class-$i$ by setting all traffic classes to zero expect class-$i$, ($\lambda_j = 0$, $j = 1 \hdots N, j \neq i$). The values of $\lambda_{d_i}$ and $C_{d_i}^A$ can be expressed as:

\begin{align} \label{eq:lambda_simple}
    \lambda_{d_i} = \lambda_d \Bigr|_{\lambda_j = 0, j \neq i; \lambda_i > 0} 
    \hspace{3mm} \mathrm{~~~and~~~} \hspace{3mm}
     C_{d_i}^A = C_{d}^A \Bigr|_{\lambda_j = 0, j \neq i; \lambda_i > 0}
\end{align}
Subsequently, we apply the principle of superposition to obtain the distribution parameters of $Q_d$ as shown in Figure~\ref{fig:superpos}(c). First, we compute $\lambda_d$ by adding all $\lambda_{d_i}$ as:
\begin{equation} \label{eq:lambda_0}
    \lambda_d = \sum_{i=1}^N \lambda_{d_i}
\end{equation}
The value of $C_{d}^A$ is approximated by applying the superposition-based traffic merging process~\cite{pujolle1986solution} for each $C_{d_i}^A$, as shown below:
\begin{equation} \label{eq:ca_0}
    (C_{d}^A)^2 = \sum_{i=1}^N \frac{\lambda_{d_i}}{\lambda_d} (C_{d_i}^A)^2    
\end{equation}
\begin{figure}[b]
\centering
\vspace{-5mm}
\includegraphics[width=0.9\linewidth]{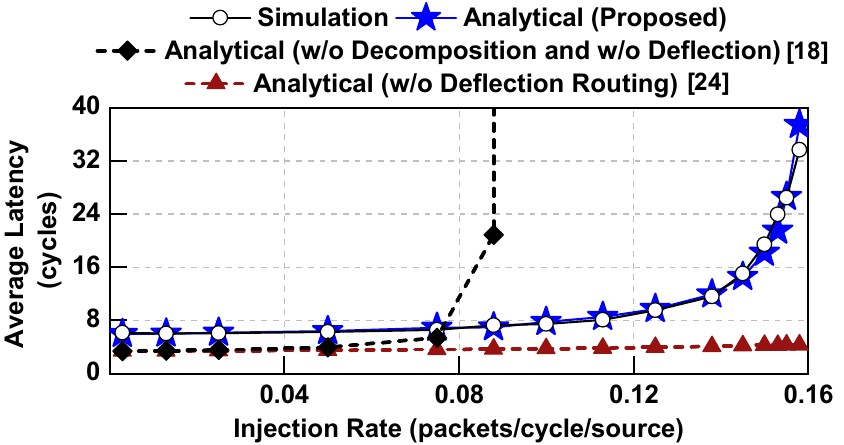}
\caption{Comparison of average latency between simulation and analytical model for the canonical example shown in Figure~\ref{fig:superpos} with $p_{d}=0.3$ and $N=5$.}
\label{fig:graph_ring_geo_canonical}
\end{figure}
Next, we use these distribution parameters ($\lambda_d, C_{d}^A$) of the deflected packets to calculate the waiting time of the traffic classes in the system. The formulation of the priority-aware queuing system is applied to obtain the waiting time of each traffic class-$i$ ($W_i$)~\cite{bertsekas1992data}:

\noindent
\begin{align} \nonumber \label{eq:mult_class_W}
    W_i = & \frac{\rho_d(T_d+1) + 2\rho_i W_d}{2(1-\rho_d - \sum_{n=1}^{i} \rho_n)} + 
    \frac{\sum_{n=1}^{i-1} (\rho_n (T_n+1) + 2\rho_i W_n)}{2(1-\rho_d - \sum_{n=1}^{i} \rho_n)} + \\
    & \frac{\rho_i(T_i-1) + T_i((C_{i}^A)^2 + \lambda_i - 1)}{2(1-\rho_d - \sum_{n=1}^{i} \rho_n)}
\end{align}
The first term in Equation~\ref{eq:mult_class_W} denotes the effect of deflected traffic on class-$i$; the second term denotes the effect of higher priority classes (class-$j$, $j<i$) on class-$i$; and the last term denotes the effect of class-$i$ itself. For more complex scenarios that include traffic splits, we apply an iterative decomposition algorithm~\cite{mandal2020analytical} to obtain the queuing time of different classes.

Figure~\ref{fig:graph_ring_geo_canonical} shows the average latency comparison between the proposed analytical model and simulation for the system in Figure~\ref{fig:superpos}. 
In this setup, we assume the number of classes is 5 ($N=5$), $p_d=0.3$, and input traffic distribution is geometric. The results show that the analytical model performs well against the simulation, with only 4\% error on average. In contrast, the analytical model from~\cite{kiasari2013analytical} highly overestimates the latency as it does not consider multiple traffic classes. The performance model of the priority-aware NoC in~\cite{mandal2019analytical} accounts for multiple traffic classes, but it does not model deflection. 
Hence, it severely underestimates the average latency. 

\begin{algorithm}[b]
\caption{End-to-end latency computation} \label{algo:latency_algo}
\SetAlgoLined
\SetNoFillComment
\textbf{Input:} NoC topology, routing algorithm, service process, input distribution for each class\camready{,} ($\lambda$, $C_A$), deflection probability ($p_d$) for each sink \\
\textbf{Output:} Average end-to-end latency ($L_{avg}$) \\
$\mathcal{S}$ = set of all classes in the network\\
$N$ = number of queues in the network\\
$\mathcal{S}_n$ = set of classes in queue $n$\\

\tcc{\textbf{Distribution of deflected traffic}}
\For {i = \normalfont{1: |$\mathcal{S}$|}} {
Compute $\lambda_{di}$ and $C_{d_i}^A$ using Equation~\ref{eq:lambda_simple} \\
Compute $\lambda_d$ and $C_{d}^A$ using Equation~\ref{eq:lambda_0} and Equation~\ref{eq:ca_0} 
}

Compute $W_d$ using $\lambda_d$ and $C_{d}^A$ \\

\tcc{\textbf{Average waiting time of each class}}

\For {$n$ = \normalfont{1:$N$}}  {
\For {$s$ = \normalfont{1:$|\mathcal{S}_n|$}}  {
Compute $W_{ns}$ using Equation~\ref{eq:mult_class_W} (if $|\mathcal{S}_n| = 1$) \\
Compute $W_{ns}$ following the decomposition method in \cite{mandal2020analytical} (if $|\mathcal{S}_n| > 1$) \\
}}
$L_{avg} = \frac{\sum_{n=1}^N \sum_{s=1}^{\mathcal{S}_n} (W_{ns} + L_{ns})\lambda_{ns}}{\sum_{n=1}^N \sum_{s=1}^{\mathcal{S}_n} \lambda_{ns}}$ (\textit{For mesh this term includes the latency both on the rows and the columns.})

\end{algorithm}


\subsection{\hspace{-3mm} Summary \& End-to-End Latency Estimation} \label{sec:algorithm}

\noindent\textbf{Summary of the analytical modeling:}
We presented a scalable approach for \camready{the} analytical model generation of end-to-end latency that handles multiple traffic classes of priority-aware NoCs with deflection routing. It applies the principle of superposition on subsystems where each subsystem is a canonical queuing system of a single traffic class to significantly simplify the approximation of the GGeo parameters of deflected traffic and in turn\camready{,} the latency calculations. 

\noindent\textbf{End-to-End latency computation:} Algorithm~\ref{algo:latency_algo} describes the end-to-end latency computation with our proposed analytical model. 
The input parameters of the algorithm are the NoC topology, routing algorithm, service process of each server, input traffic distribution for each class\camready{,} and deflection probability per sink. It outputs the average end-to-end latency ($L_{avg}$).
First, the queuing system is decomposed into multiple subsystems as shown in Figure~\ref{fig:superpos}(b) and $\lambda_{d_i}$ and $C_{d_i}^A$ for each subsystem-$i$ are computed.
Subsequently, the proposed superposition methodology is applied to compute $\lambda_d$ and $C_{d}^A$, shown in lines 6--9 of the algorithm.
Then, $\lambda_d$ and $C_{d}^{A}$ are used to compute the average waiting time of the deflected packets ($W_d$).
Then, the average waiting time for class-$s$ in $Q_n$ ($W_{ns}$) is computed as shown in lines 13--14.
The service time combined with static latency from source to destination ($L_{ns}$) is added to $W_{ns}$ to obtain the end-to-end latency.
Finally, the average end-to-end latency ($L_{avg}$) is computed by taking a weighted average of the latency of each class\camready{,} as shown in line 16 of the algorithm.




\section{Experimental Evaluations} \label{sec:expt_eval}

This section validates the proposed analytical model against an industrial cycle-accurate NoC simulator under a wide range of traffic scenarios.
The experiment scenarios \camready{include} real applications and synthetic traffic that \camready{allow} evaluations with varying injection rates and deflection probabilities. 
The evaluations include a 6$\times$6 mesh NoC and a 6$\times$1 ring as representative examples of 
high-end server CPUs~\cite{jeffers2016intel} and 
high-end client CPUs~\cite{rotem2015intel}, respectively. 
In both cases, the traffic sources emulate high-end CPU cores with a 100\% hit rate on the shared last level cache (LLC) to load the NoCs. 
\camready{The} target platforms are more powerful than experimental~\cite{vangal200880} and special-purpose~\cite{wentzlaff2007chip} platforms with simple cores, although the mesh size is smaller. 
To further demonstrate the scalability of the proposed approach, we also present results with mesh sizes up to 16$\times$16. 
All cycle-accurate simulations run for 200K cycles, with a warm-up period of 20K cycles, to allow the NoC to reach the steady-state.



\begin{figure}[b]
\centering
 \includegraphics[width=0.9\linewidth]{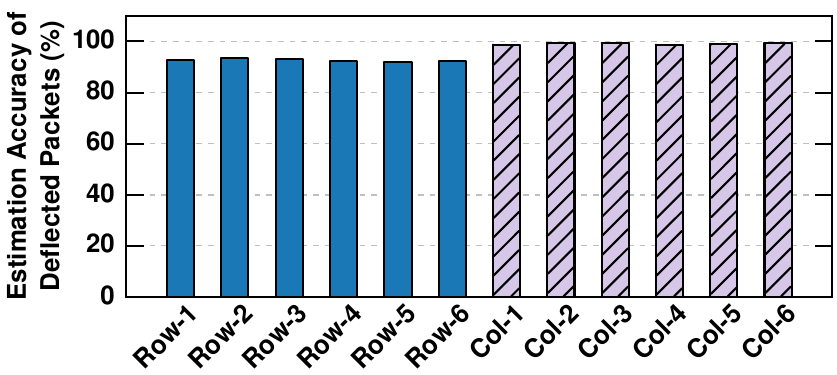}
\vspace{-4mm}
\caption{Estimation accuracy of average number of packets deflected for each row and column in a 6$\times$6 mesh with $p_d$=0.3.}
\vspace{-4mm}
\label{fig:lambda_bnc_accuracy}
\end{figure}


\begin{figure*}[t]
\centering
\vspace{-15mm}
\includegraphics[width=0.9\linewidth]{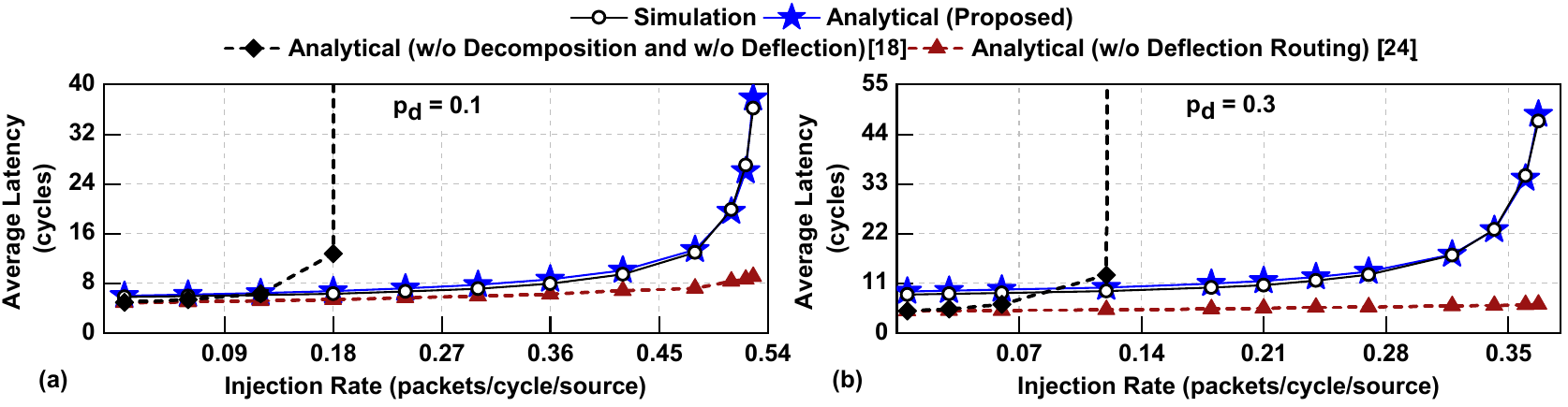}
\caption{Comparison of average latency between simulation, the analytical model proposed in this work, and analytical models proposed in~\cite{kiasari2013analytical, mandal2019analytical} for a 6$\times$6 mesh with deflection probability (a) 0.1 and (b) 0.3.}
\label{fig:graph_mesh_geo}
\end{figure*}

\begin{table*}[b!]
\setlength\tabcolsep{.4pt}
\centering
\vspace{-1mm}
\captionsetup{justification=centering}
\caption{Validation of the proposed analytical model for 6$\times$6 mesh and 6$\times$1 ring \\ with bursty traffic arrival, and comparisons against prior work~\cite{kiasari2013analytical,mandal2019analytical}. `E' signifies error >100\%.}
\vspace{-3mm}
\label{tab:ring_mesh_bursty_validation}
\begin{tabular}{|c|c||c|c|c||c|c|c||c|c|c||c|c|c||c|c|c||c|c|c||c|c|c||c|c|c||c|c|c||c|c|c||c|c|c||c|c|c|}
\hline
\multicolumn{2}{|c||}{\textbf{Topo.}} & \multicolumn{18}{c||}{\textbf{6$\times$6 Mesh}} & \multicolumn{18}{c|}{\textbf{6$\times$1 Ring}} \\ \hline
\multicolumn{2}{|c||}{$p_{d}$} & \multicolumn{6}{c||}{0.1} & \multicolumn{6}{c||}{0.2} & \multicolumn{6}{c||}{0.3} & \multicolumn{6}{c||}{0.1} & \multicolumn{6}{c||}{0.2} & \multicolumn{6}{c|}{0.3} \\ \hline
\multicolumn{2}{|c||}{$p_{br}$} & \multicolumn{3}{c||}{0.2} & \multicolumn{3}{c||}{0.6} & \multicolumn{3}{c||}{0.2} & \multicolumn{3}{c||}{0.6} & \multicolumn{3}{c||}{0.2} & \multicolumn{3}{c||}{0.6} & \multicolumn{3}{c||}{0.2} & \multicolumn{3}{c||}{0.6} & \multicolumn{3}{c||}{0.2} & \multicolumn{3}{c||}{0.6} & \multicolumn{3}{c||}{0.2} & \multicolumn{3}{c|}{0.6} \\ \hline
\multicolumn{2}{|c||}{$\lambda$} & 0.1 & 0.3 & 0.4 & 0.1 & 0.3 & 0.4 & 0.1 & 0.3 & 0.4 & 0.1 & 0.3 & 0.4 & 0.1 & 0.2 & 0.3 & 0.1 & 0.2 & 0.3 & 0.1 & 0.3 & 0.4 & 0.1 & 0.3 & 0.4 & 0.1 & 0.3 & 0.4 & 0.1 & 0.3 & 0.4 & 0.1 & 0.2 & 0.3 & 0.1 & 0.2 & 0.3 \\ \hline
\multirow{3}{*}{\rotatebox{90}{Error(\%)}} & \textbf{Prop.} & \textbf{7.3} & \textbf{9.6} & \textbf{8.1} & \textbf{14} & \textbf{13} & \textbf{14} & \textbf{8.9} & \textbf{8.0} & \textbf{7.7} & \textbf{13} & \textbf{12} & \textbf{12} & \textbf{9.6} & \textbf{9.2} & \textbf{6.5} & \textbf{11} & \textbf{12} & \textbf{13} & \textbf{1.0} & \textbf{4.1} & \textbf{5.8} & \textbf{4.6} & \textbf{5.2} & \textbf{5.5} & \textbf{0.7} & \textbf{2.3} & \textbf{4.2} & \textbf{6.3} & \textbf{7.3} & \textbf{8.6} & \textbf{0.7} & \textbf{0.9} & \textbf{3.3} & \textbf{6.3} & \textbf{8.5} & \textbf{8.6} \\ \cline{2-38} 
 & {\cellcolor{Gray}} Ref{}\cite{kiasari2013analytical}{} & {\cellcolor{Gray}}2.6 & {\cellcolor{Gray}}E & {\cellcolor{Gray}}E & {\cellcolor{Gray}}26 & {\cellcolor{Gray}}E & {\cellcolor{Gray}}E & {\cellcolor{Gray}}22 & {\cellcolor{Gray}}E & {\cellcolor{Gray}}E & {\cellcolor{Gray}}39 & {\cellcolor{Gray}}E & {\cellcolor{Gray}}E & {\cellcolor{Gray}}35 & {\cellcolor{Gray}}18 & {\cellcolor{Gray}}E & {\cellcolor{Gray}}57 & {\cellcolor{Gray}}E &{\cellcolor{Gray}}E & {\cellcolor{Gray}}7.0 & {\cellcolor{Gray}}E & {\cellcolor{Gray}}E & {\cellcolor{Gray}}34 & {\cellcolor{Gray}}E & {\cellcolor{Gray}}E & {\cellcolor{Gray}}23 & {\cellcolor{Gray}}E & {\cellcolor{Gray}}E & {\cellcolor{Gray}}45 & {\cellcolor{Gray}}E & {\cellcolor{Gray}}E & {\cellcolor{Gray}}42 & {\cellcolor{Gray}}E & {\cellcolor{Gray}}E & {\cellcolor{Gray}}54 & {\cellcolor{Gray}}E & {\cellcolor{Gray}}E \\ \cline{2-38} 
  & {\cellcolor{Maroon}} Ref\cite{mandal2019analytical}{} & {\cellcolor{Maroon}}12 & {\cellcolor{Maroon}}15 & {\cellcolor{Maroon}}23 & {\cellcolor{Maroon}}3.1 & {\cellcolor{Maroon}}18 & {\cellcolor{Maroon}}23 & {\cellcolor{Maroon}}28 & {\cellcolor{Maroon}}41 & {\cellcolor{Maroon}}65 & {\cellcolor{Maroon}}19 & {\cellcolor{Maroon}}33 & {\cellcolor{Maroon}}49 & {\cellcolor{Maroon}}42 & {\cellcolor{Maroon}}45 & {\cellcolor{Maroon}}55 & {\cellcolor{Maroon}}39 & {\cellcolor{Maroon}}35 & {\cellcolor{Maroon}}31 & {\cellcolor{Maroon}}15.3 & {\cellcolor{Maroon}}18 & {\cellcolor{Maroon}}22 & {\cellcolor{Maroon}}18 & {\cellcolor{Maroon}}24 & {\cellcolor{Maroon}}33 & {\cellcolor{Maroon}}30 & {\cellcolor{Maroon}}38 & {\cellcolor{Maroon}}67 & {\cellcolor{Maroon}}31 & {\cellcolor{Maroon}}44 & {\cellcolor{Maroon}}54 & {\cellcolor{Maroon}}41 & {\cellcolor{Maroon}}50 & {\cellcolor{Maroon}}73 & {\cellcolor{Maroon}}42 & {\cellcolor{Maroon}}50 & {\cellcolor{Maroon}}58 \\ \hline
\end{tabular}
\end{table*}



\subsection{Estimation of Deflected Traffic}
One of the key components of the proposed analytical model is estimating the average number of deflected packets.
This section evaluates the accuracy of this estimation compared to simulation with a 6$\times$6 mesh.
To perform evaluations under heavy load, we set the deflection probability at each junction and sink to $p_d=0.3$ and injection rates at each source to 0.33 packets/cycle/source, which are relatively large values seen in actual systems.
We first run cycle-accurate simulations to obtain the average number of deflected packets at each row and column of the mesh.
Then, the analytical model estimates the same quantities for the 6$\times$6 mesh.
Figure~\ref{fig:lambda_bnc_accuracy} shows the estimation accuracy for all rows and columns.
The average estimation accuracy across all rows and columns is 96\% and the worst-case accuracy is 92\%.
Overall, this evaluation shows that the proposed model accurately estimates the average number of deflected packets.



\subsection{\hspace{-3mm} Evaluations with Geometric Traffic Input}
\label{sec:geometric}


This section evaluates the accuracy of our latency estimation technique when the sources inject packets following a geometric traffic distribution. 
We note that our technique can also handle bursty traffic, which is significantly harder. 
However, we start with this assumption to make a fair comparison to two state-of-the-art techniques from the literature~\cite{kiasari2013analytical,mandal2019analytical}.
The model presented in~\cite{kiasari2013analytical} does not incorporate multiple traffic classes and deflection routing.
On the other hand, the model presented in~\cite{mandal2019analytical} considers multiple traffic classes but does not consider bursty traffic and deflection routing.

The evaluations are performed first on the server-like 6$\times$6 mesh for deflection probabilities $p_{d}=$~0.1 and $p_{d}=$~0.3 while sweeping the packet injection rates.
Figure~\ref{fig:graph_mesh_geo}(a) and Figure~\ref{fig:graph_mesh_geo}(b) show that the proposed technique follows the simulation results closely for all injections. 
More specifically, the proposed analytical model has only 7\% and 6\% percentage error on average for deflection probabilities of 0.1 and 0.3\camready{,} respectively. 
In 
\camready{sharp}
contrast, the analytical model proposed in~\cite{kiasari2013analytical} significantly overestimates the latency 
starting with moderate injection rates, 
since it does not consider multiple traffic classes. 
Its performance degrades even further with larger deflection probability, as depicted in Figure~\ref{fig:graph_mesh_geo}(b). 
We note that it also slightly underestimates the latency at low injection rates since it ignores deflection.
Unlike this approach, the technique presented in~\cite{mandal2019analytical} 
considers multiple traffic classes in the same queue, but it ignores deflected packets. 
Consequently, it severely underestimates the latency impact of deflection, as shown in Figure~\ref{fig:graph_mesh_geo}. 

We repeated the same evaluation on a 6$\times$1 priority-aware ring NoC which follows a high-end industrial quad-core CPU with an integrated GPU and memory-controller~\cite{jeffers2016intel}.
The average error between the proposed analytical model and simulations are 7\% and 4\% for
\camready{deflection} probabilities of 0.1 and 0.3, respectively. 
In contrast, the model presented in~\cite{kiasari2013analytical} underestimates the latency at low injection rates and significantly overestimates it under high traffic load similar to the 6$\times$6 results in Figure~\ref{fig:graph_mesh_geo}. 
Similarly, the analytical model presented in~\cite{mandal2019analytical} severely underestimates the average latency. 
It leads to an average 43\% error with respect to simulation.
The plots of these results are not included for space considerations since they closely follow the results in Figure~\ref{fig:graph_mesh_geo}.

\begin{figure*}[t]
\centering
\vspace{-8mm}
 \includegraphics[width=0.9\linewidth]{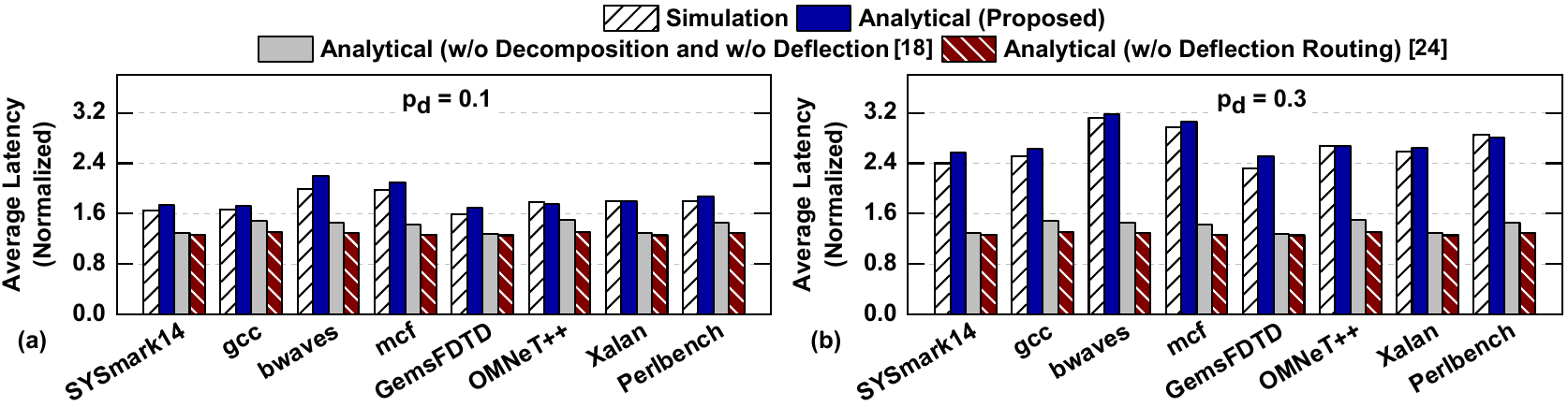}
 \vspace{-3mm}
\caption{Average latency comparison between simulation, the analytical model proposed in this work, and analytical models proposed in~\cite{kiasari2013analytical, mandal2019analytical} for a 6$\times$6 mesh with (a) $p_{d}$ = 0.1 and (b) $p_{d}$ = 0.3.}
\vspace{-2mm}
\label{fig:graph_mesh_real_apps}
\end{figure*}

\subsection{\hspace{-3mm}Latency Estimation with Bursty Traffic Input}
\label{sec:bursty_validation}

Since real applications exhibit burstiness, it is crucial to perform accurate analytical modeling under bursty traffic.
Therefore, this section presents the comparison of our proposed analytical model with respect to simulation under bursty traffic.
For an extensive and thorough validation, we sweep the packet injection rate ($\lambda$), probability of burstiness ($p_{br}$), and deflection probability ($p_{d}$).
The injection rates cover a wide range to capture various traffic congestion scenarios in the network.
Likewise, we report evaluation results for two different burstiness 
($p_{br}=$~\{0.2,0.6\}), 
and three different deflection probabilities ($p_d=$~\{0.1,0.2,0.3\}). 
The coefficient of variation for the input traffic ($C_{A}$), 
the final input to the model, 
is then computed as a function of $p_{br}$ and $\lambda$~\cite{kouvatsos1984analysis}. We simulate the 6$\times$6 mesh and 6$\times$1 ring NoCs using their cycle-accurate models for all input parameter values mentioned above. 
Then, we estimate the average packet latencies using the proposed technique, 
as well as the most relevant prior work~\cite{kiasari2013analytical,mandal2019analytical}.

The estimation error of all three performance analysis techniques is reported  in Table~\ref{tab:ring_mesh_bursty_validation} for all input parameters. 
The mean and median estimation errors of our proposed technique are 9.3\% and 9.5\%, respectively. Furthermore, we do not observe more than 14\% error even with relatively higher traffic load, probability of deflection, and burstiness than seen in real applications (presented in the following section).
In strong contrast, the analytical models proposed in~\cite{kiasari2013analytical} severely overestimate the latency similar to the results presented in Section~\ref{sec:geometric}. 
The estimation error is more than 100\% for most cases since the impact of multiple traffic classes and deflected packet become more significant under these challenging scenarios. 
Similarly, the model proposed in~\cite{mandal2019analytical} underestimates the latency because it does not model bursty traffic. 

The right-hand side of Table~\ref{tab:ring_mesh_bursty_validation} summarizes the estimation errors obtained on the 6$\times$1 ring NoC that follows high-end client systems. 
In most cases, the error with the proposed analytical model is within 10\% of simulation, and the error is as low as 1\%.
With $p_{d}=0.1$, $p_{br}=0.6$ and $\lambda=0.4$, the error is 14\%, which is acceptable, considering that the network is severely congested.
In contrast, the analytical models proposed in~\cite{kiasari2013analytical} overestimate the latency, whereas the models in~\cite{mandal2019analytical} underestimate the latency which conforms the results with geometric traffic, 
as in the 6$\times$6 mesh results. 



\vspace{-3mm}
\subsection{Experiments with Real Applications}
\label{sec:real_apps}

In addition to the synthetic traffic, the proposed analytical model is evaluated with applications from SPEC CPU\textregistered 2006~\cite{henning2006spec}, SPEC CPU\textregistered 2017 benchmark suites~\cite{bucek2018spec}, and the SYSmark\textregistered 2014 application~\cite{SYSmark2014}.
Specifically, the evaluation includes SYSmark14, gcc, bwaves, mcf, GemsFDTD, OMNeT++, Xalan, and perlbench applications.
The chosen applications represent a variety of injection rates for each source in the NoC and different levels of burstiness.
Each application is profiled offline to find the input traffic parameters. Of note, the probability of burstiness
for these applications ranges from $p_b=$~0.25 to $p_b=$~0.55, 
which is aligned with the evaluations in Section~\ref{sec:bursty_validation}.

The benchmark applications are executed on both 6$\times$6 mesh and 6$\times$1 ring architectures.
The comparison of average latency between simulation and proposed analytical model for the 6$\times$6 mesh is shown in  Figure~\ref{fig:graph_mesh_real_apps}. 
The proposed model follows the simulation results very closely for deflection probability $p_{d}=$~0.1 and $p_{d}=$~0.3, 
as shown in Figure~\ref{fig:graph_mesh_real_apps}(a) and Figure~\ref{fig:graph_mesh_real_apps}(b), respectively. 
These plots show the average packet latencies normalized to the smallest latency from the 6$\times$1 ring simulations due to \camready{the} confidentiality of the results. 
On average, the proposed analytical model achieves less than 5\% modeling error. 
In contrast, the analytical models which do not consider deflection routing~\cite{kiasari2013analytical, mandal2019analytical} underestimate the latency, 
since the injection rates of these applications are in the range of 0.02--0.1 flits/cycle/source (low injection region). 

We observe similar results for the 6$\times$1 ring NoC. 
The average estimation error of our proposed technique is less than 8\% for all applications. 
In contrast, the prior techniques underestimate the latency by more than 2$\times$ since they ignore deflected packets, and the average traffic loads are small. 
In conclusion, the proposed technique outperforms state-of-the-art for real applications and a wide range of synthetic traffic inputs. 


\begin{figure}[t]
\centering
\vspace{-3mm}
\includegraphics[width=0.88\linewidth]{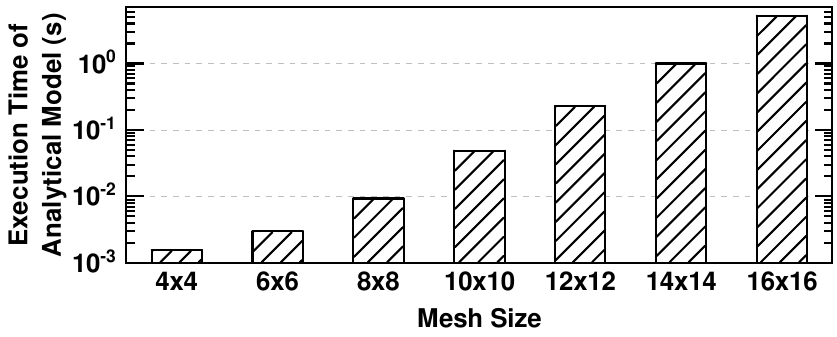}
\vspace{-4mm}
\caption{Execution time of the proposed analytical model (in seconds) for different mesh sizes.}
\vspace{-3mm}
\label{fig:graph_scalability}
\end{figure}

\subsection{Scalability Analysis}
Finally, we evaluate the scalability of the proposed technique for larger NoCs. We note that accuracy results for larger NoCs are not available since they do not have detailed cycle-accurate simulation models. 
We implemented the analytical model in C.
Figure~\ref{fig:graph_scalability} shows that the analysis completes in the order of seconds for up to 16$\times$16 mesh. 
In comparison, cycle-accurate simulations 
\camready{take hours}
with this size, even considering linear scaling.
When we scale the mesh size aggressively to 16$\times$16, the analysis completes in about 5 seconds, which is orders of magnitude faster than cycle-accurate simulations of NoCs of this size.

\section{Conclusions} \label{sec:concl}


Industrial NoCs incorporate priority-arbitration and deflection routing to minimize buffer requirement and achieve predictable latency.
Analytical performance modeling of these NoCs is needed to perform design space exploration, fast system simulations, and tuning architectural parameters.
However, state-of-the-art performance analysis models for NoCs do not incorporate priority arbitration and deflection routing together.
This paper presented a performance analysis technique for industrial priority aware NoCs with deflection routing under heavy traffic.
Experimental evaluations with industrial NoCs show that the proposed technique significantly outperforms existing analytical models under both real application and a wide range of synthetic traffic workloads. 

\bibliographystyle{ACM-Reference-Format}
\bibliography{main.bbl}


\begin{thebibliography}{36}


\ifx \showCODEN    \undefined \def \showCODEN     #1{\unskip}     \fi
\ifx \showDOI      \undefined \def \showDOI       #1{#1}\fi
\ifx \showISBNx    \undefined \def \showISBNx     #1{\unskip}     \fi
\ifx \showISBNxiii \undefined \def \showISBNxiii  #1{\unskip}     \fi
\ifx \showISSN     \undefined \def \showISSN      #1{\unskip}     \fi
\ifx \showLCCN     \undefined \def \showLCCN      #1{\unskip}     \fi
\ifx \shownote     \undefined \def \shownote      #1{#1}          \fi
\ifx \showarticletitle \undefined \def \showarticletitle #1{#1}   \fi
\ifx \showURL      \undefined \def \showURL       {\relax}        \fi
\providecommand\bibfield[2]{#2}
\providecommand\bibinfo[2]{#2}
\providecommand\natexlab[1]{#1}
\providecommand\showeprint[2][]{arXiv:#2}

\bibitem[\protect\citeauthoryear{Agarwal, Krishna, Peh, and Jha}{Agarwal
  et~al\mbox{.}}{2009}]%
        {agarwal2009garnet}
\bibfield{author}{\bibinfo{person}{Niket Agarwal}, \bibinfo{person}{Tushar
  Krishna}, \bibinfo{person}{Li-Shiuan Peh}, {and} \bibinfo{person}{Niraj~K
  Jha}.} \bibinfo{year}{2009}\natexlab{}.
\newblock \showarticletitle{{GARNET: A Detailed on-chip Network Model inside a
  Full-System Simulator}}. In \bibinfo{booktitle}{\emph{2009 IEEE International
  Symposium on Performance Analysis of Systems and Software}}.
  \bibinfo{pages}{33--42}.
\newblock


\bibitem[\protect\citeauthoryear{(BAPCo)}{(BAPCo)}{[n.d.]}]%
        {SYSmark2014}
\bibfield{author}{\bibinfo{person}{Business Applications
  Performance~Corporation (BAPCo)}.} \bibinfo{year}{[n.d.]}\natexlab{}.
\newblock \bibinfo{booktitle}{\emph{Benchmark, SYSmark2014}}.
\newblock
\newblock
\shownote{\url{http://bapco.com/products/sysmark-2014}, accessed 27 May 2020.}


\bibitem[\protect\citeauthoryear{Bertsekas, Gallager, and Humblet}{Bertsekas
  et~al\mbox{.}}{1992}]%
        {bertsekas1992data}
\bibfield{author}{\bibinfo{person}{Dimitri~P Bertsekas},
  \bibinfo{person}{Robert~G Gallager}, {and} \bibinfo{person}{Pierre Humblet}.}
  \bibinfo{year}{1992}\natexlab{}.
\newblock \bibinfo{booktitle}{\emph{{Data Networks}}}.
  Vol.~\bibinfo{volume}{2}.
\newblock \bibinfo{publisher}{Prentice-Hall International New Jersey}.
\newblock


\bibitem[\protect\citeauthoryear{Binkert et~al\mbox{.}}{Binkert
  et~al\mbox{.}}{2011}]%
        {Binkert2011Gem5}
\bibfield{author}{\bibinfo{person}{Nathan Binkert} {et~al\mbox{.}}}
  \bibinfo{year}{2011}\natexlab{}.
\newblock \showarticletitle{{The Gem5 Simulator}}.
\newblock \bibinfo{journal}{\emph{SIGARCH Computer Architecture News}}
  (\bibinfo{date}{May.} \bibinfo{year}{2011}).
\newblock


\bibitem[\protect\citeauthoryear{Bogdan and Marculescu}{Bogdan and
  Marculescu}{2010}]%
        {bogdan2010workload}
\bibfield{author}{\bibinfo{person}{Paul Bogdan} {and} \bibinfo{person}{Radu
  Marculescu}.} \bibinfo{year}{2010}\natexlab{}.
\newblock \showarticletitle{{Workload Characterization and Its Impact on
  Multicore Platform Design}}. In \bibinfo{booktitle}{\emph{2010 IEEE/ACM/IFIP
  International Conference on Hardware/Software Codesign and System Synthesis
  (CODES+ ISSS)}}. \bibinfo{pages}{231--240}.
\newblock


\bibitem[\protect\citeauthoryear{Bolch, Greiner, De~Meer, and Trivedi}{Bolch
  et~al\mbox{.}}{2006}]%
        {bolch2006queueing}
\bibfield{author}{\bibinfo{person}{Gunter Bolch}, \bibinfo{person}{Stefan
  Greiner}, \bibinfo{person}{Hermann De~Meer}, {and} \bibinfo{person}{Kishor~S
  Trivedi}.} \bibinfo{year}{2006}\natexlab{}.
\newblock \bibinfo{booktitle}{\emph{{Queueing Networks and Markov Chains:
  Modeling and Performance Evaluation with Computer Science Applications}}}.
\newblock \bibinfo{publisher}{John Wiley \& Sons}.
\newblock


\bibitem[\protect\citeauthoryear{Borodin, Rabani, and Schieber}{Borodin
  et~al\mbox{.}}{1997}]%
        {borodin1997deterministic}
\bibfield{author}{\bibinfo{person}{Allan Borodin}, \bibinfo{person}{Yuval
  Rabani}, {and} \bibinfo{person}{Baruch Schieber}.}
  \bibinfo{year}{1997}\natexlab{}.
\newblock \showarticletitle{{Deterministic Many-to-Many Hot Potato Routing}}.
\newblock \bibinfo{journal}{\emph{IEEE Transactions on Parallel and Distributed
  Systems}} \bibinfo{volume}{8}, \bibinfo{number}{6} (\bibinfo{year}{1997}),
  \bibinfo{pages}{587--596}.
\newblock


\bibitem[\protect\citeauthoryear{Brassil and Cruz}{Brassil and Cruz}{1995}]%
        {brassil1995bounds}
\bibfield{author}{\bibinfo{person}{Jack~T Brassil} {and}
  \bibinfo{person}{Rene~L. Cruz}.} \bibinfo{year}{1995}\natexlab{}.
\newblock \showarticletitle{{Bounds on Maximum Delay in Networks with
  Deflection Routing}}.
\newblock \bibinfo{journal}{\emph{IEEE Transactions on Parallel and Distributed
  Systems}} \bibinfo{volume}{6}, \bibinfo{number}{7} (\bibinfo{year}{1995}),
  \bibinfo{pages}{724--732}.
\newblock


\bibitem[\protect\citeauthoryear{Bucek, Lange, and v.~Kistowski}{Bucek
  et~al\mbox{.}}{2018}]%
        {bucek2018spec}
\bibfield{author}{\bibinfo{person}{James Bucek}, \bibinfo{person}{Klaus-Dieter
  Lange}, {and} \bibinfo{person}{J{\'o}akim v. Kistowski}.}
  \bibinfo{year}{2018}\natexlab{}.
\newblock \showarticletitle{{SPEC CPU2017: Next-Generation Compute Benchmark}}.
  In \bibinfo{booktitle}{\emph{Companion of the 2018 ACM/SPEC International
  Conference on Performance Engineering}}. \bibinfo{pages}{41--42}.
\newblock


\bibitem[\protect\citeauthoryear{Fallin, Craik, and Mutlu}{Fallin
  et~al\mbox{.}}{2011}]%
        {fallin2011chipper}
\bibfield{author}{\bibinfo{person}{Chris Fallin}, \bibinfo{person}{Chris
  Craik}, {and} \bibinfo{person}{Onur Mutlu}.} \bibinfo{year}{2011}\natexlab{}.
\newblock \showarticletitle{{CHIPPER: A Low-Complexity Bufferless Deflection
  Router}}. In \bibinfo{booktitle}{\emph{2011 IEEE 17th International Symposium
  on High Performance Computer Architecture}}. \bibinfo{pages}{144--155}.
\newblock


\bibitem[\protect\citeauthoryear{Fallin, Nazario, Yu, Chang, Ausavarungnirun,
  and Mutlu}{Fallin et~al\mbox{.}}{2012}]%
        {fallin2012minbd}
\bibfield{author}{\bibinfo{person}{Chris Fallin}, \bibinfo{person}{Greg
  Nazario}, \bibinfo{person}{Xiangyao Yu}, \bibinfo{person}{Kevin Chang},
  \bibinfo{person}{Rachata Ausavarungnirun}, {and} \bibinfo{person}{Onur
  Mutlu}.} \bibinfo{year}{2012}\natexlab{}.
\newblock \showarticletitle{{MinBD: Minimally-buffered Deflection Routing for
  Energy-efficient Interconnect}}. In \bibinfo{booktitle}{\emph{2012 IEEE/ACM
  Sixth International Symposium on Networks-on-Chip}}. \bibinfo{pages}{1--10}.
\newblock


\bibitem[\protect\citeauthoryear{Ghosh and Givargis}{Ghosh and
  Givargis}{2003}]%
        {ghosh2003analytical}
\bibfield{author}{\bibinfo{person}{Arijit Ghosh} {and} \bibinfo{person}{Tony
  Givargis}.} \bibinfo{year}{2003}\natexlab{}.
\newblock \showarticletitle{{Analytical Design Space Exploration of Caches for
  Embedded Systems}}. In \bibinfo{booktitle}{\emph{2003 Design, Automation and
  Test in Europe Conference and Exhibition}}. \bibinfo{pages}{650--655}.
\newblock


\bibitem[\protect\citeauthoryear{Ghosh, Ravi, and Sen}{Ghosh
  et~al\mbox{.}}{2010}]%
        {ghosh2010analytical}
\bibfield{author}{\bibinfo{person}{Pavel Ghosh}, \bibinfo{person}{Arvind Ravi},
  {and} \bibinfo{person}{Arunabha Sen}.} \bibinfo{year}{2010}\natexlab{}.
\newblock \showarticletitle{{An Analytical Framework with Bounded Deflection
  Adaptive Routing for Networks-on-Chip}}. In \bibinfo{booktitle}{\emph{2010
  IEEE Computer Society Annual Symposium on VLSI}}. \bibinfo{pages}{363--368}.
\newblock


\bibitem[\protect\citeauthoryear{Henning}{Henning}{2006}]%
        {henning2006spec}
\bibfield{author}{\bibinfo{person}{John~L Henning}.}
  \bibinfo{year}{2006}\natexlab{}.
\newblock \showarticletitle{{SPEC CPU2006 Benchmark Descriptions}}.
\newblock \bibinfo{journal}{\emph{ACM SIGARCH Computer Architecture News}}
  \bibinfo{volume}{34}, \bibinfo{number}{4} (\bibinfo{year}{2006}),
  \bibinfo{pages}{1--17}.
\newblock


\bibitem[\protect\citeauthoryear{Jeffers, Reinders, and Sodani}{Jeffers
  et~al\mbox{.}}{2016}]%
        {jeffers2016intel}
\bibfield{author}{\bibinfo{person}{James Jeffers}, \bibinfo{person}{James
  Reinders}, {and} \bibinfo{person}{Avinash Sodani}.}
  \bibinfo{year}{2016}\natexlab{}.
\newblock \bibinfo{booktitle}{\emph{{Intel Xeon Phi Processor High Performance
  Programming: Knights Landing Edition}}}.
\newblock \bibinfo{publisher}{Morgan Kaufmann}.
\newblock


\bibitem[\protect\citeauthoryear{Jiang et~al\mbox{.}}{Jiang
  et~al\mbox{.}}{[n.d.]}]%
        {jiang2013detailed}
\bibfield{author}{\bibinfo{person}{Nan Jiang} {et~al\mbox{.}}}
  \bibinfo{year}{[n.d.]}\natexlab{}.
\newblock \showarticletitle{{A Detailed and Flexible Cycle-accurate
  Network-on-chip Simulator}}. In \bibinfo{booktitle}{\emph{2013 IEEE Intl.
  Symp. on Performance Analysis of Systems and Software (ISPASS)}}.
  \bibinfo{pages}{86--96}.
\newblock


\bibitem[\protect\citeauthoryear{Kashif and Patel}{Kashif and Patel}{2014}]%
        {kashif2014bounding}
\bibfield{author}{\bibinfo{person}{Hany Kashif} {and} \bibinfo{person}{Hiren
  Patel}.} \bibinfo{year}{2014}\natexlab{}.
\newblock \showarticletitle{{Bounding Buffer Space Requirements for Real-time
  Priority-aware Networks}}. In \bibinfo{booktitle}{\emph{2014 19th Asia and
  South Pacific Design Automation Conference (ASP-DAC)}}.
  \bibinfo{pages}{113--118}.
\newblock


\bibitem[\protect\citeauthoryear{Kiasari, Lu, and Jantsch}{Kiasari
  et~al\mbox{.}}{2013}]%
        {kiasari2013analytical}
\bibfield{author}{\bibinfo{person}{Abbas~Eslami Kiasari},
  \bibinfo{person}{Zhonghai Lu}, {and} \bibinfo{person}{Axel Jantsch}.}
  \bibinfo{year}{2013}\natexlab{}.
\newblock \showarticletitle{{An Analytical Latency Model for
  Networks-on-Chip}}.
\newblock \bibinfo{journal}{\emph{IEEE Transactions on Very Large Scale
  Integration (VLSI) Systems}} \bibinfo{volume}{21}, \bibinfo{number}{1}
  (\bibinfo{year}{2013}), \bibinfo{pages}{113--123}.
\newblock


\bibitem[\protect\citeauthoryear{Kouvatsos and Luker}{Kouvatsos and
  Luker}{1984}]%
        {kouvatsos1984analysis}
\bibfield{author}{\bibinfo{person}{DD Kouvatsos} {and} \bibinfo{person}{PA
  Luker}.} \bibinfo{year}{1984}\natexlab{}.
\newblock \showarticletitle{ON THE ANALYSIS OF QUEUEING NETWORK MODELS: MAXIMUM
  ENTROPY AND SIMULATION}.
\newblock In \bibinfo{booktitle}{\emph{UKSC 84}}. \bibinfo{pages}{488--496}.
\newblock


\bibitem[\protect\citeauthoryear{Kouvatsos}{Kouvatsos}{1994}]%
        {kouvatsos1994entropy}
\bibfield{author}{\bibinfo{person}{Demetres~D Kouvatsos}.}
  \bibinfo{year}{1994}\natexlab{}.
\newblock \showarticletitle{{Entropy Maximisation and Queuing Network Models}}.
\newblock \bibinfo{journal}{\emph{Annals of Operations Research}}
  \bibinfo{volume}{48}, \bibinfo{number}{1} (\bibinfo{year}{1994}),
  \bibinfo{pages}{63--126}.
\newblock


\bibitem[\protect\citeauthoryear{Leupers, Eeckhout, Martin, Schirrmeister,
  Topham, and Chen}{Leupers et~al\mbox{.}}{2011}]%
        {leupers2011virtual}
\bibfield{author}{\bibinfo{person}{Rainer Leupers}, \bibinfo{person}{Lieven
  Eeckhout}, \bibinfo{person}{Grant Martin}, \bibinfo{person}{Frank
  Schirrmeister}, \bibinfo{person}{Nigel Topham}, {and}
  \bibinfo{person}{Xiaotao Chen}.} \bibinfo{year}{2011}\natexlab{}.
\newblock \showarticletitle{{Virtual Manycore Platforms: Moving Towards 100+
  Processor Cores}}. In \bibinfo{booktitle}{\emph{2011 Design, Automation \&
  Test in Europe}}. IEEE, \bibinfo{pages}{1--6}.
\newblock


\bibitem[\protect\citeauthoryear{Lu, Zhong, and Jantsch}{Lu
  et~al\mbox{.}}{2006}]%
        {lu2006evaluation}
\bibfield{author}{\bibinfo{person}{Zhonghai Lu}, \bibinfo{person}{Mingchen
  Zhong}, {and} \bibinfo{person}{Axel Jantsch}.}
  \bibinfo{year}{2006}\natexlab{}.
\newblock \showarticletitle{{Evaluation of On-chip Networks using Deflection
  Routing}}. In \bibinfo{booktitle}{\emph{Proceedings of the 16th ACM Great
  Lakes Symposium on VLSI}}. \bibinfo{pages}{296--301}.
\newblock


\bibitem[\protect\citeauthoryear{Mandal, Ayoub, Kishinevsky, Islam, and
  Ogras}{Mandal et~al\mbox{.}}{2020}]%
        {mandal2020analytical}
\bibfield{author}{\bibinfo{person}{Sumit~K Mandal}, \bibinfo{person}{Raid
  Ayoub}, \bibinfo{person}{Michael Kishinevsky}, \bibinfo{person}{Mohammad~M
  Islam}, {and} \bibinfo{person}{Umit~Y Ogras}.}
  \bibinfo{year}{2020}\natexlab{}.
\newblock \showarticletitle{{Analytical Performance Modeling of NoCs under
  Priority Arbitration and Bursty Traffic}}.
\newblock \bibinfo{journal}{\emph{IEEE Embedded Systems Letters}}
  (\bibinfo{year}{2020}).
\newblock


\bibitem[\protect\citeauthoryear{Mandal, Ayoub, Kishinevsky, and Ogras}{Mandal
  et~al\mbox{.}}{2019}]%
        {mandal2019analytical}
\bibfield{author}{\bibinfo{person}{Sumit~K Mandal}, \bibinfo{person}{Raid
  Ayoub}, \bibinfo{person}{Michael Kishinevsky}, {and} \bibinfo{person}{Umit~Y
  Ogras}.} \bibinfo{year}{2019}\natexlab{}.
\newblock \showarticletitle{{Analytical Performance Models for NoCs with
  Multiple Priority Traffic Classes}}.
\newblock \bibinfo{journal}{\emph{ACM Transactions on Embedded Computing
  Systems (TECS)}} \bibinfo{volume}{18}, \bibinfo{number}{5s}
  (\bibinfo{year}{2019}).
\newblock


\bibitem[\protect\citeauthoryear{Moscibroda and Mutlu}{Moscibroda and
  Mutlu}{2009}]%
        {moscibroda2009case}
\bibfield{author}{\bibinfo{person}{Thomas Moscibroda} {and}
  \bibinfo{person}{Onur Mutlu}.} \bibinfo{year}{2009}\natexlab{}.
\newblock \showarticletitle{{A Case for Bufferless Routing in On-chip
  Networks}}. In \bibinfo{booktitle}{\emph{Proceedings of the 36th Annual
  International Symposium on Computer Architecture}}.
  \bibinfo{pages}{196--207}.
\newblock


\bibitem[\protect\citeauthoryear{Ogras, Bogdan, and Marculescu}{Ogras
  et~al\mbox{.}}{2010}]%
        {ogras2010analytical}
\bibfield{author}{\bibinfo{person}{Umit~Y Ogras}, \bibinfo{person}{Paul
  Bogdan}, {and} \bibinfo{person}{Radu Marculescu}.}
  \bibinfo{year}{2010}\natexlab{}.
\newblock \showarticletitle{{An Analytical Approach for Network-on-Chip
  Performance Analysis}}.
\newblock \bibinfo{journal}{\emph{IEEE Transactions on Computer-Aided Design of
  Integrated Circuits and Systems}} \bibinfo{volume}{29}, \bibinfo{number}{12}
  (\bibinfo{year}{2010}), \bibinfo{pages}{2001--2013}.
\newblock


\bibitem[\protect\citeauthoryear{Palesi and Givargis}{Palesi and
  Givargis}{2002}]%
        {palesi2002multi}
\bibfield{author}{\bibinfo{person}{Maurizio Palesi} {and} \bibinfo{person}{Tony
  Givargis}.} \bibinfo{year}{2002}\natexlab{}.
\newblock \showarticletitle{{Multi-objective Design Space Exploration using
  Genetic Algorithms}}. In \bibinfo{booktitle}{\emph{Proceedings of the tenth
  international symposium on Hardware/software codesign}}.
  \bibinfo{pages}{67--72}.
\newblock


\bibitem[\protect\citeauthoryear{Petracca, Lee, Bergman, and Carloni}{Petracca
  et~al\mbox{.}}{2009}]%
        {petracca2009photonic}
\bibfield{author}{\bibinfo{person}{Michele Petracca},
  \bibinfo{person}{Benjamin~G Lee}, \bibinfo{person}{Keren Bergman}, {and}
  \bibinfo{person}{Luca~P Carloni}.} \bibinfo{year}{2009}\natexlab{}.
\newblock \showarticletitle{{Photonic NoCs: System-Level Design Exploration}}.
\newblock \bibinfo{journal}{\emph{IEEE Micro}} \bibinfo{volume}{29},
  \bibinfo{number}{4} (\bibinfo{year}{2009}), \bibinfo{pages}{74--85}.
\newblock


\bibitem[\protect\citeauthoryear{Pujolle and Ai}{Pujolle and Ai}{1986}]%
        {pujolle1986solution}
\bibfield{author}{\bibinfo{person}{Guy Pujolle} {and} \bibinfo{person}{Wu Ai}.}
  \bibinfo{year}{1986}\natexlab{}.
\newblock \showarticletitle{{A Solution for Multiserver and Multiclass Open
  Queueing Networks}}.
\newblock \bibinfo{journal}{\emph{INFOR: Information Systems and Operational
  Research}} \bibinfo{volume}{24}, \bibinfo{number}{3} (\bibinfo{year}{1986}),
  \bibinfo{pages}{221--230}.
\newblock


\bibitem[\protect\citeauthoryear{Qian, Juan, Bogdan, Tsui, Marculescu, and
  Marculescu}{Qian et~al\mbox{.}}{2015}]%
        {qian2015support}
\bibfield{author}{\bibinfo{person}{Zhi-Liang Qian}, \bibinfo{person}{Da-Cheng
  Juan}, \bibinfo{person}{Paul Bogdan}, \bibinfo{person}{Chi-Ying Tsui},
  \bibinfo{person}{Diana Marculescu}, {and} \bibinfo{person}{Radu Marculescu}.}
  \bibinfo{year}{2015}\natexlab{}.
\newblock \showarticletitle{{A Support Vector Regression (SVR)-based Latency
  Model for Network-on-Chip (NoC) Architectures}}.
\newblock \bibinfo{journal}{\emph{IEEE Transactions on Computer-Aided Design of
  Integrated Circuits and Systems}} \bibinfo{volume}{35}, \bibinfo{number}{3}
  (\bibinfo{year}{2015}), \bibinfo{pages}{471--484}.
\newblock


\bibitem[\protect\citeauthoryear{Rotem}{Rotem}{2015}]%
        {rotem2015intel}
\bibfield{author}{\bibinfo{person}{Efraim Rotem}.}
  \bibinfo{year}{2015}\natexlab{}.
\newblock \showarticletitle{{Intel Architecture, Code Name Skylake Deep Dive: A
  New Architecture to Manage Power Performance and Energy Efficiency}}. In
  \bibinfo{booktitle}{\emph{Intel Developer Forum}}.
\newblock


\bibitem[\protect\citeauthoryear{Vangal et~al\mbox{.}}{Vangal
  et~al\mbox{.}}{2008}]%
        {vangal200880}
\bibfield{author}{\bibinfo{person}{Sriram~R Vangal} {et~al\mbox{.}}}
  \bibinfo{year}{2008}\natexlab{}.
\newblock \showarticletitle{{An 80-tile sub-100-w teraflops processor in 65-nm
  cmos}}.
\newblock \bibinfo{journal}{\emph{IEEE Journal of Solid-State Circuits}}
  \bibinfo{volume}{43}, \bibinfo{number}{1} (\bibinfo{year}{2008}),
  \bibinfo{pages}{29--41}.
\newblock


\bibitem[\protect\citeauthoryear{Walraevens}{Walraevens}{2004}]%
        {walraevens2004discrete}
\bibfield{author}{\bibinfo{person}{Joris Walraevens}.}
  \bibinfo{year}{2004}\natexlab{}.
\newblock \emph{\bibinfo{title}{{Discrete-time Queueing Models with
  Priorities}}}.
\newblock \bibinfo{thesistype}{Ph.D. Dissertation}. \bibinfo{school}{Ghent
  University}.
\newblock


\bibitem[\protect\citeauthoryear{Wentzlaff et~al\mbox{.}}{Wentzlaff
  et~al\mbox{.}}{2007}]%
        {wentzlaff2007chip}
\bibfield{author}{\bibinfo{person}{David Wentzlaff} {et~al\mbox{.}}}
  \bibinfo{year}{2007}\natexlab{}.
\newblock \showarticletitle{{On-chip Interconnection Architecture of the Tile
  Processor}}.
\newblock \bibinfo{journal}{\emph{IEEE micro}} \bibinfo{volume}{27},
  \bibinfo{number}{5} (\bibinfo{year}{2007}), \bibinfo{pages}{15--31}.
\newblock


\bibitem[\protect\citeauthoryear{Wu, Wu, Min, Ould-Khaoua, Yin, and Wang}{Wu
  et~al\mbox{.}}{2010}]%
        {wu2010analytical}
\bibfield{author}{\bibinfo{person}{Yulei Wu}, \bibinfo{person}{Yulei Wu},
  \bibinfo{person}{Geyong Min}, \bibinfo{person}{Mohamed Ould-Khaoua},
  \bibinfo{person}{Hao Yin}, {and} \bibinfo{person}{Lan Wang}.}
  \bibinfo{year}{2010}\natexlab{}.
\newblock \showarticletitle{{Analytical Modelling of Networks in Multicomputer
  Systems under Bursty and Batch Arrival Traffic}}.
\newblock \bibinfo{journal}{\emph{The Journal of Supercomputing}}
  \bibinfo{volume}{51}, \bibinfo{number}{2} (\bibinfo{year}{2010}),
  \bibinfo{pages}{115--130}.
\newblock


\bibitem[\protect\citeauthoryear{Yoo, Nicolescu, Gauthier, and Jerraya}{Yoo
  et~al\mbox{.}}{2002}]%
        {yoo2002automatic}
\bibfield{author}{\bibinfo{person}{Sungjoo Yoo}, \bibinfo{person}{Gabriela
  Nicolescu}, \bibinfo{person}{Lovic Gauthier}, {and}
  \bibinfo{person}{Ahmed~Amine Jerraya}.} \bibinfo{year}{2002}\natexlab{}.
\newblock \showarticletitle{{Automatic Generation of Fast Timed Simulation
  Models for Operating Systems in SoC Design}}. In
  \bibinfo{booktitle}{\emph{Proceedings 2002 Design, Automation and Test in
  Europe Conference and Exhibition}}. \bibinfo{pages}{620--627}.
\newblock


\end{thebibliography}

\end{document}